# Routing Attacks in Wireless Sensor Networks: A Survey

Deepali Virmani[#1], Ankita Soni[*2], Shringarica Chandel[*2], Manas Hemrajani[*2]

*Bhagwan Parshuram Institute of Technology, India*

[1]deepalivirmani@gmail.com
[2]ankita24soni@gmail.com
[3]shringaricachandel37@gmail.com
[4]manas.hemrajani@gmail.com

*Abstract*— **Wireless Sensor Networks (WSN) is an emerging technology now-a-days and has a wide range of applications such as battlefield surveillance, traffic surveillance, forest fire detection, flood detection etc. But wireless sensor networks are susceptible to a variety of potential attacks which obstructs the normal operation of the network. The security of a wireless sensor network is compromised because of the random deployment of sensor nodes in open environment, memory limitations, power limitations and unattended nature. This paper focuses on various attacks that manifest in the network and provides a tabular representation of the attacks, their effects and severity. The paper depicts a comparison of attacks basis packet loss and packet corruption. Also, the paper discusses the known defence mechanisms and countermeasures against the attacks.**

*Keywords*— **wireless sensor network, security, attacks, defence mechanism.**

## I. INTRODUCTION

Wireless Sensor Network consists of a large number of small and low cost sensor nodes which are randomly deployed in an area. The sensor nodes have computational capability to carry out simple computations and transmit the required information [1]. These nodes transmit information to the sink node that aggregates the entire information received from other nodes and generates a summary data to be transmitted to another network. These sensor nodes can collectively monitor physical and environmental conditions like pressure, temperature, humidity and sound vibrations. Such features ensure a wide range of applications for wireless sensor network such as military, medical, industrial, disaster relief operations, environmental monitoring, traffic surveillance, agriculture, infrastructure monitoring [1, 2]. Since the majority of sensor nodes are deployed in hostile environment, they are susceptible to various attacks that are caused by malicious or compromised nodes in the network. The malicious nodes can alter the normal behaviour of the network, tamper with the node's hardware and software, transmit false information, or drop the required information. Hence, security of wireless sensor network becomes a critical issue.

This paper majorly contributes towards the security attacks and their defence mechanisms. The paper is structured in the following manner: Section 2 discusses about the security goals in wireless sensor network. Section 3 provides a categorization of security attacks. Section 4 gives their detailed explanation. Section 5 gives an explanation of the known defence mechanisms to counter the attacks. Section 6 depicts a graphical representation of the comparison of attacks. Section 7 concludes the paper.

## II. SECURITY GOALS IN WIRELESS SENSOR NETWORKS

A wireless sensor network shares some common features with the traditional network and also has unique features of its own that distinguishes it from the traditional network. Therefore, the security goals or requirements cover both the traditional network goals and the goals suited solely to the wireless sensor network. The security goals can be classified into two types: Primary and Secondary goals. We discuss both of them.

### A. Primary Goals

1) *Data Confidentiality:* Confidentiality is the means of limiting information access to only the authorized users and preventing access or disclosure by the unauthorized users. Data confidentiality is the most important issue that any network must address. Sensor nodes carry sensitive data which must be concealed from the malicious nodes or attackers. If sensor nodes are not capable of keeping the data confidential, then any neighbouring node can tamper with the data and transmit false information. This can cause serious hazards, especially in military applications.

2) *Data Authentication:* Data authentication is the ability of a receiver to verify that the data received by it is from a correct sender. In a wireless sensor network, data can not only be tampered by the malicious nodes but the entire packet stream can be changed by addition of false packets to it. So, a receiver must be able to identify if the data originated from the correct source or not. Data authentication can be achieved using symmetric key cryptography where the sender and receiver share a secret key or using asymmetric key cryptography where the data can be encrypted and decrypted using public and private keys.

3) *Data Availability:* Data availability determines if the services of the network are available in case of failure or presence of attacks in the network. A single point failure in the network can threaten the availability of resources and other services. So, data availability is of prime importance and is responsible for the operation of the network.

*4) Data Integrity:* The malicious nodes in the network can manipulate the data in the packets [3]. Data integrity ensures that the received data is not altered in transit. It confirms that the data is reliable and has not been altered or changed. The network must incorporate security mechanisms against different attacks caused by malicious nodes so as to ensure integrity of the data.

## B. Secondary Goals

*1) Data Freshness:* Data freshness determines that the data is recent and no old packets have been replayed. It is important to ensure the freshness of the message, apart from ensuring data confidentiality and integrity. There are two types of data freshness: Weak freshness that provides partial message ordering but doesn't provide any delay information and strong freshness, which provides total message ordering and delay estimation [4]. Weak freshness is used for sensor measurements while strong freshness is employed in time synchronization in the network.

*2) Self-Organization:* The sensor nodes in a wireless sensor network are randomly deployed and have no fixed infrastructure. So, these sensor nodes must have self-organising capability so that they can dynamically organise according to the environment and situation. Self-organising capability is important to ensure multi-hop routing, key management and building trust relations with the neighbours. If self-organising capability lacks in a sensor network, then damage resulting from attacks can be significant.

*3) Time Synchronisation:* Most sensor network applications rely on some form of time synchronization. When a packet travels between two pairwise sensors, sensors can compute the end-to-end delay of a packet. A more collaborative sensor network may require group synchronization for tracking applications [5].

*4) Secure Organization:* The utility of a sensor network relies on its ability to accurately and automatically locate each sensor in the network. Wireless sensor networks which are expected to locate faults needs accurate information about a location in order to indicate a fault's location. Unfortunately, a malicious node can manipulate non secured location information by reporting false signal strengths, replaying signals [6].

## III. ATTACKS IN WIRELESS SENSOR NETWORKS

A passive attack involves monitoring and listening of the data stream but doesn't involve modification of the data stream. Passive attacks do not cause direct harm to the network as they cannot modify the data. Attacks against privacy are a passive attack. The classification of passive attacks is shown in Fig.1.

## A. Attacks against privacy

Sensor networks allow the availability of large volumes of data through remote access. This causes a privacy problem as the malicious nodes can easily obtain information without being physically available to maintain surveillance. So, the malicious nodes can gather information at low-risk in an anonymous manner [7]. The attacks against privacy are classified into three categories as shown in Fig.1.

*1) Eavesdropping:* In eavesdropping, a malicious node simply overhears the data stream to gain knowledge about the communication content. When the network traffic transmits control information about the sensor network configuration that contains detailed information about the network, eavesdropping can act effectively against privacy protection.

*2) Traffic Analysis:* Malicious nodes can analyse the network traffic to determine which nodes have high activity. Once the highly active sensor nodes are discovered, the malicious nodes can cause harm to those sensor nodes.

*3) Camouflage:* Malicious nodes can hide in the sensor network by masquerading as normal sensor nodes. So they deceive the other sensor nodes and attract packets from them. After receiving the packets, malicious nodes can either misroute the packets or eventually drop the packets.

## B. Active Attacks

An active attack involves monitoring, listening and modification of the data stream by the malicious nodes/adversaries prevailing inside or outside the network. Active attacks cause direct harm to the network because they can manipulate the data stream. There are many types of active attacks. In this paper we focus mainly on the routing attacks in the network. The classification of active attacks is shown in Fig.1.

## IV. ROUTING ATTACKS IN WIRELESS SENSOR NETWORKS

The attacks which act on the network layer are called routing attacks. These attacks occur while routing messages. We discuss the following routing attacks.

## A. Sybil Attack

Sybil Attack is named after the subject of the book *Sybil*, a case study of a woman diagnosed with multiple fake identities. These fake identities are known as Sybil nodes. The Sybil nodes can out vote the honest nodes in the system. Usually, peer to peer systems are vulnerable to Sybil attack. Examples of vulnerable systems include vehicular Ad hoc Network, Distributed Storage Applications in Peer to Peer Systems, Routing in a Distributed Peer to Peer System [8], etc.

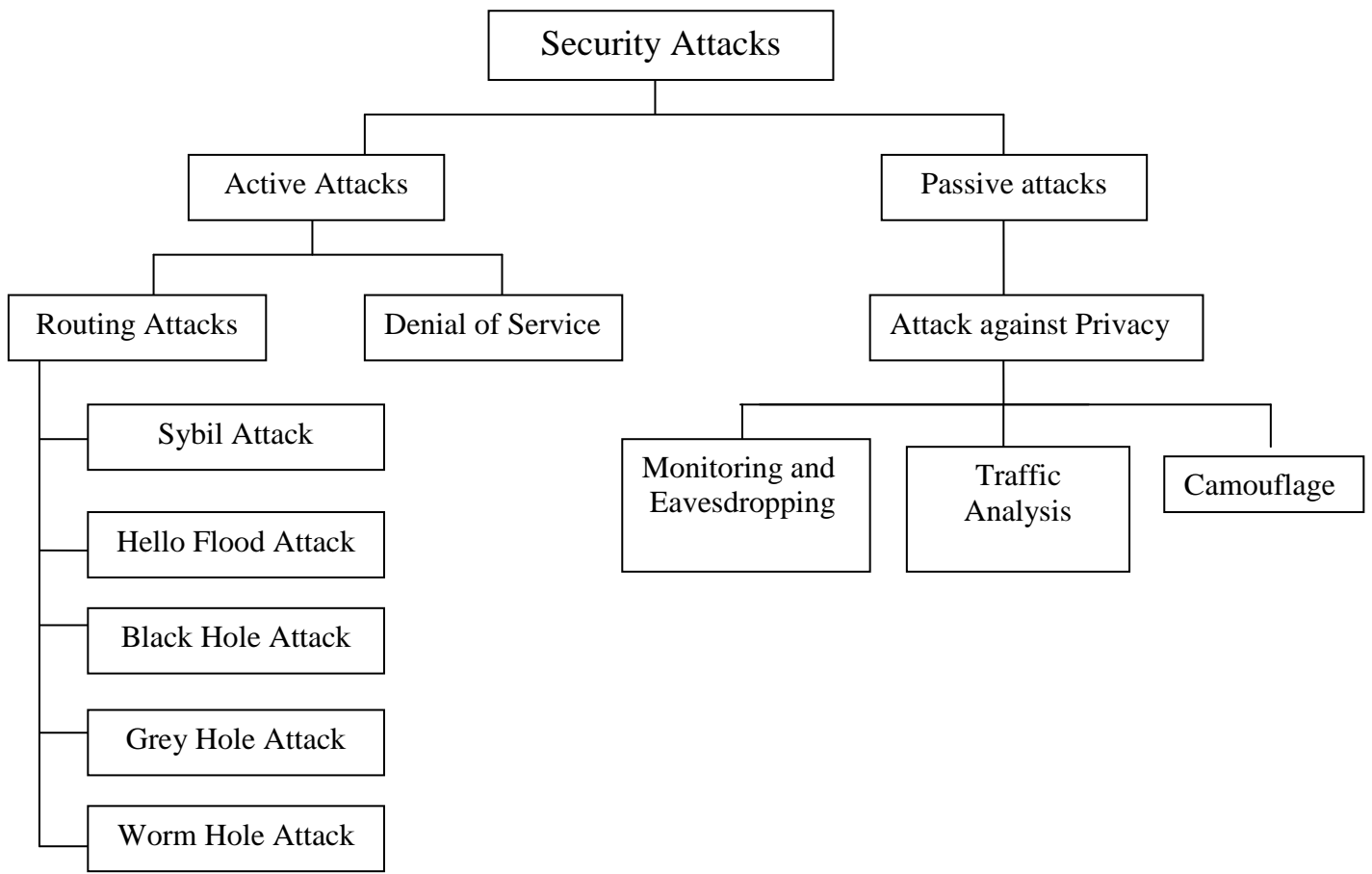

Fig. 1. Classification of Attacks on WSN.

B. *Black hole Attack*

A black hole is a malicious node that attracts all the traffic in the network by advertising that it has the shortest path in the network [9]. So, it creates a metaphorical black hole with the malicious node or the adversary at the center. This black hole drops all the packets it receives from the other nodes.

In a black hole attack, malicious nodes do not send true control messages. To execute a black hole attack, malicious node awaits the neighboring nodes to send RREQ messages. When the malicious nodes receive RREQ message from its neighboring nodes, it immediately sends a false RREP message providing a route to the destination over itself. In this way, it assigns a high sequence number to be settled in the routing table of victim node, before true nodes send a genuine reply. Therefore, requesting nodes assume that route discovery process is completed and ignore RREP messages from other nodes and start sending packets over malicious node. This is how malicious nodes attacks all the RREQ messages and takes over all the routes in the network. Therefore, all the packets are sent to the malicious node from where they are never forwarded and eventually dropped. This is called a black hole akin to real meaning which swallows all objects and matter [10].

C. *Denial of Service Attack*

A Denial of Service (DoS) attack is one that attempts to prevent the victim from being able to use all or part of his/her network connection [11]. DoS attack allows an adversary to subvert, disrupt, or destroy a network, and also to diminish a network's capability to provide a service [5].

Dos attack extends to all the layers of the protocol stack. They are usually very difficult to prevent because they exist in many forms inside the network. For example, a malicious node can send huge number of requests to a server which has to test the legitimacy of the nodes. Due to the huge number of requests, the server will be busy in testing illegal requests and so, it will not be available for the legal users. This causes performance degradation of the entire network as the network gets congested because of illegal requests.

D. *Wormhole Attack*

Wormhole attack is an attack on the routing protocol in which the packets or individual bits of the packets are captured at one location, tunnelled to another location and then replayed at another location [12],[13]. In this attack the two colluding nodes create an illusion that the locations involved are directly connected though they are actually distant.

E. *Hello Flood Attack*

Most protocols require nodes to broadcast HELLO PACKETS to show their presence to their neighbours and the receiving nodes may assume that it is within the RF range of the sender. This assumption may prove to be false when a laptop-class attacker transmit routing information with an abnormally high transmission power to prove every other node in the network that the malicious node is its neighbour. Such an attack in the network is called a hello flood attack [14].

F. *Grey hole Attack*

A grey hole attack is a variation of black hole attack in which the nodes selectively drops packets [15]. There are two ways in which a node can drop packets:
- It can drop all UDP packets while transmitting all TCP packets.
- It can drop 50% of the packets or can drop them with probabilistic distribution.

In a grey hole attack a normal node can prevent from behaving usually and therefore this attack is difficult to detect. A grey hole attack affects one or two nodes in the network whereas a black hole attack affects the whole network.

V. COUNTERMEASURES AGAINST ROUTING ATTACKS

A. *Countermeasure against Sybil Attack*

*1) Trusted certification:* This type of method assumes that there is a special trusted third party or central authority, which can verify the validity of each participant, and further issues a certification for the honest one [8]. In reality, such certification can be a special hardware device [16] or a digital number [17]. Note that essentially both of them are a series of digits present on different medias. Before a participant joins a peer-to-peer system to provide votes or to obtain its services, his identity must first be verified [18].

*2) Sybil Guard:* The Sybil nodes in a Sybil attack are connected to the honest nodes via attack edges. Attack edges are difficult for Sybil nodes to create and hence they are few in number. It leads to Sybil nodes and honest nodes being completely isolated and connected together by a few attack edges. From a trusted node, there are a number of random paths with fixed length known as verifiers. The Sybil guard checks a suspected node by sending random paths from the suspected node. If the random path intersects with verifier then the suspected node is said to be verified once. After the node is verified a particular number of times, the suspected node is said to be trusted node otherwise it is said to be a Sybil node [19].

B. *Countermeasures against Black hole Attack*

*1) REWARD technique:* REWARD is a routing technique where a wireless sensor network is organized as a distributed data base to detect black hole attack. The distributed data base maintains a record for suspicious nodes and areas. This routing algorithm consists of two types of broadcast messages, MISS (material for intersection of suspicious sets) and SAMBA (suspicious area, mark a black-hole attack). The destination node broadcasts a MISS message when it doesn't receive a packet within a specified time. The destination copies the list of all the involved nodes to the MISS message. The nodes listed in the MISS message are counted as suspicious nodes. The SAMBA message provides the location of the black-hole attack. If a malicious node does not forward packets, the previous node in the path will broadcast a SAMBA message [20].

*2) Path based Detection Algorithm:* In path based approach, a node watches only the next hop neighbor in the current route path rather than observing every node in the neighbor [21]. To implement the algorithm, every node maintains a FwdPktBuffer (packet digest buffer). When a packet is forwarded, its digest it added to the FwdPktBuffer and the detecting node overhears the transmission. Once it is overheard that the next hop forwarded the packet, the digest is released from the FwdPktBuffer. The detecting node calculates the overhear rate of its next hop neighbor and compares it with the threshold. If the forwarding rate is lower than the threshold value, the detecting node considers the next hop neighbor as a black hole and avoids sending packets via the suspect node in future.

$$OR(N) = \frac{\text{total overheard packet number}}{\text{total forwarded packet number}}$$

*3) Exponential Trust based mechanism:* In this method, a streak counter (n) is maintained which keeps track of the packets that have been dropped consecutively. If a packet is dropped the counter is incremented but if a packet is forwarded the counter is reset to zero. It uses the fact that in a black hole attack all the packets are dropped. A tolerance factor (X) is set for the environment in which the mechanism is deployed. The mechanism uses the streak counter to calculate a trust factor using the formula $100*x^n$ for each node [22]. When a packet is dropped the trust factor drops exponentially. When the trust factor goes below a threshold value the node is declared as malicious.

*4) Reliability Analysis mechanism:* This method combines AODV protocol with reliability analysis [23]. It consists of a DRI table which keeps track of the no. of packets sent and received. Based on this information, it calculates the reliability ratio of the route that consists of the neighbors of node.

$$Reliability\ Ratio = \frac{No.of\ Packets\ Sent}{No.of\ Packets\ Received}$$

It also consists of an REL packet which is sent when the reliable route has been discovered. REL packets maintain the count of reliability for each node.

*C. Countermeasures against Denial of Service Attack*

*1) Firewall and Router Filtering:* A firewall is a router or a computer that monitors packet traffic and protects the system from malicious access. Firewalls can be used as a relay or as a semi-transparent gateway for DoS countermeasure [11].

*Firewall as Semi-transparent gateway:* The firewall sends SYN packets to the host and the host replies with a SYN+ACK packet. When the firewall receives SYN+ACK packet from the host, it forwards it to the client and also sends an ACK packet to the host. If the firewall does not receive an ACK from the client within a certain timeout period, it terminates the connection by sending an RST packet to the host. The duplicate ACK that arrives at the host is discarded by the TCP protocol for legitimate connections, and succeeding packets flow without intervention by the firewall [24]. In this approach, no delays are introduced for legitimate connections.

*Firewall as a Relay:* The firewall responds on behalf of the internal host. During an attack, the firewall responds to the SYN sent by the attacker; since the ACK never arrives, an RST packet is sent to terminate the connection, and the host never receives the datagram. For legal connections, the firewall creates a new connection to the internal host on behalf of the client, and continues to perform as a proxy for translation of sequence numbers of packets that flow between the client and server [24]. In this approach, the host is completely shielded against a DoS attack and doesn't receive spoofed SYN packets ever.

*2) Cookies:* Cookie-based approach uses a one-way hash function to verify if the connection requests are legitimate. Bernstein and Bona suggested a stateless cookie approach. When a client sends a SYN packet, the server calculates a one-way hash of the sender's sequence number, ports, the server's secret key, and a counter which changes after every one minute [25]. The server sends the result of the one-way hash to the client. When the client replies with an ACK packet, the server again calculates the same hash function and discards the packet if it fails to authenticate with the server [25]. Otherwise, the server sets up the connection, if it doesn't already exist. Since CPU time is used to calculate hash values, memory is never exhausted by SYN flood DOS attacks.

*D. Countermeasures against Wormhole Attack*

*1) Packet Leashes:* A leash is an extra piece of information that is added to a packet to restrict its maximum travel distance. There are two types of leashes: geographical leashes and temporal leashes. A geographical leash [12] ensures that the recipient of the packet is within a certain distance from the sender [12]. A temporal leash makes sure that the packet has a certain upper bound on its lifetime, which restricts its maximum travel distance. Both the types of leashes can be used to prevent wormhole attack.

*2) True Link:* True Link is a countermeasure which protects against wormhole attack using the combination of two phases: rendezvous phase and authentication phase. True link considers two nodes i and j. In the rendezvous phase, i and j exchange randomly generated numbers known as a nonce [26]. This phase is completed as a single RTS-CTS-DATA-ACK exchange. In this phase the timing constraints are very strict and makes it impossible for the attacker to relay these frames as only the direct neighbour can respond in time. In the authentication phase, i and j authenticate themselves as the originator of their respective nonce by sending signed messages

*E. Countermeasures against Hello Flood Attack*

*1) Identity Verification Protocol:* One of the defence against hello flood attack involves every node to authenticate each of its neighbours with an identity verification protocol using a trusted base stations [14]. In this case hello flood attack is prevented only when the malicious node has a powerful transmitter because the protocol checks the bidirectionality of the link. This bidirectionality check does not prevent any compromised node from authenticating itself to a large number of nodes in the network. Since such a malicious node is required to authenticate itself to each and every victim before launching an attack, a malicious node claiming to be a neighbour of abnormally large number of nodes is said to be alarming.

*2) AODV-HFDP:* Another method to prevent hello flood attack is a signal strength and time and threshold based AODV-HFDP(Ad-hoc On demand Distance Routing with Hello flood Detection cum Prevention) [27].In this approach it is assumed that all nodes have same signal strength within the same radio range. Each node verifies the signal strength of the received hello packet with its own. If it is found to be the same then the node is declared as a friend else a stranger.

*F. Countermeasures against Grey hole Attack*

*1) Multipath routing:* Multipath routing can be used to protect against selective forwarding. [28]Messages routed over n completely different paths are completely protected against selective forwarding attacks involving at most n compromised nodes and still offer some probabilistic protection when over n nodes are compromised. The use of multiple braided paths (which may have nodes in common but no links in common) may provide probabilistic protection against selective forwarding using only localised information. Allowing nodes to dynamically choose a packet's next hop probabilistically from a set of possible candidates can further reduce the chances of an adversary gaining complete control of a data flow.

*2) CHEMAS (Checkpoint-based Multi-hop Acknowledgement Scheme):* This scheme uses three types of packets: event packet, ACK packets and alert packets [29]. This scheme is based on checkpoint-by-checkpoint acknowledgement instead of hop-by-hop acknowledgement. The basic idea of this scheme is based on checkpoint nodes which are selected from the part of intermediate nodes. The path is divided into several segments which consist of forwarding path between two checkpoint nodes. When the source node detects a special event it generates an event packet. The packet traverses hop-by-hop towards the base station and each intermediate node saves the event packet in its cache before sending it downstream. When the checkpoint nodes receive the event packet it generates an ACK packet and sends it to upstream neighbour. The ACK packets traverse the same but reversed path upstream. It traverses at least two segments before being dropped by an upstream checkpoint. Thus all the intermediate nodes in these two checkpoints know that previous event has safely arrived in the downstream checkpoint. If the ACK packet is not received from downstream by all the nodes in these two segments then the next downstream neighbouring node is declared as suspicious and the alert packet is generated.

## VI. COMPARISON OF ATTACKS

In this paper, we compare all the six routing attacks based on parameters like number of packets corrupted and number of packets lost. This comparison gives us an analysis of which attack can cause maximum harm to the system and decrease the reliability and security of the system. Fig. 2 depicts a comparison of attacks that clearly shows the percentage of packet loss by each attack. Fig. 3 depicts another comparison that shows the percentage of packet corruption caused by each attack.

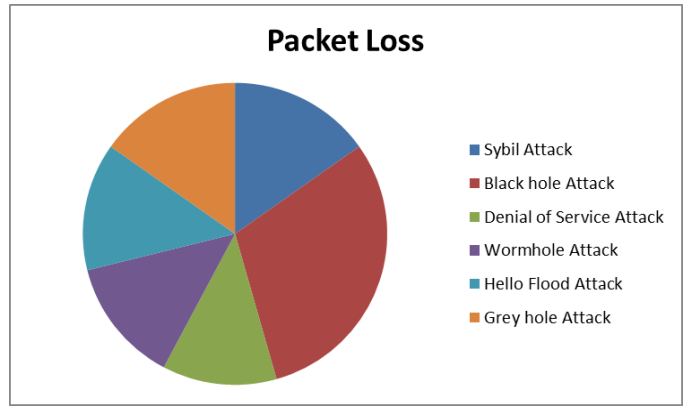

Fig. 2. Comparison of attacks based on Packet Loss.

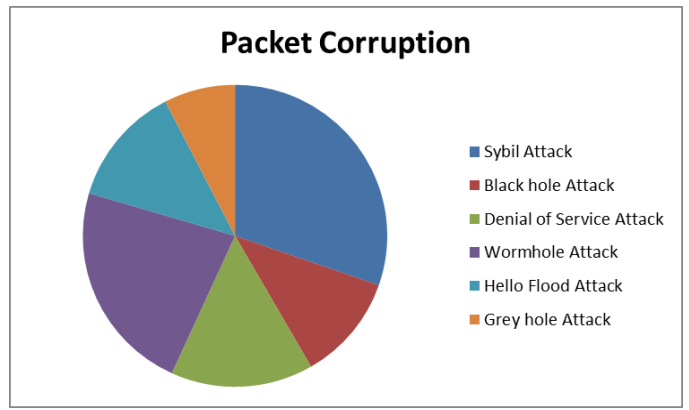

Fig. 3. Comparison of attacks based on Packet Corruption.

## VII. CONCLUSION

Wireless Sensor Networks are vulnerable to many types of attacks due to deployment of sensor nodes in an unattended environment. In this survey, firstly we have given the security goals of a network. Next, we have classified the attacks in WSN in two categories i.e. active and passive attacks. Further, we have given the definition of these types of attacks and have also given the known defences and countermeasures against them. This survey also gives the tabular classification of attacks and determines the severity of each attack. We hope that this survey will help future researches in developing a good knowledge about the attacks and their countermeasures.

TABLE 1
ATTACKS ON WSN

| Attack Name | Attack Definition | Attack Effects | Severity |
| --- | --- | --- | --- |
| Eavesdropping | Overhearing the communication channel to gather confidential data . | <ul><li>Reduces data confidentiality</li><li>Extracts vital WSN information</li><li>Threatens privacy protection of WSN</li></ul> | Low |
| Traffic Analysis | Monitoring the network traffic and computing parameters that affect the network. | <ul><li>Degradation of network performance</li><li>Increased packet collision</li><li>Increased contention</li><li>Traffic distortion</li></ul> | Low |
| Camouflage | Malicious nodes masquerade as normal nodes to attract packets. | <ul><li>Increased packet loss/corruption</li><li>False data to network</li></ul> | Low |
| Sybil | Impersonation by malicious nodes as multiple fake identities to attract packets from nodes. | <ul><li>Packet loss/ corruption</li><li>False sensor readings</li><li>Modification of routing information</li></ul> | High |
| Black hole | Attracting all the possible traffic to a compromised node. Can result in launch of other attacks. | <ul><li>Triggers other attacks like wormhole, eavesdropping</li><li>Exhausts all the network resources</li><li>Packet dropping/ corruption</li><li>Modification of routing information</li></ul> | High |
| Denial of Service (DoS) | Prevents the user from being able to use the network services. Extends to all the layers of protocol stack. | <ul><li>Reduces WSN availability</li><li>Affects physical layer, link layer, network layer, transport layer and application layer</li><li>Prevents access to network services by the user.</li></ul> | High |
| Wormhole | Tunneling and replaying messages from one location to another via low latency links that connect two nodes of WSN. | <ul><li>Changes normal message stream</li><li>False routes / misdirection</li><li>Forged routing</li><li>Changes network topology</li></ul> | High |
| Hello Flood | Transmission of a message by malicious node with an abnormally high transmission power to make the nodes believe that it is their neighbor | <ul><li>False / misleading routes generated</li><li>Route disruption</li><li>Packet loss</li><li>Confusion</li></ul> | High |
| Grey hole | Selective dropping of packets by attracting packets to a compromised node. | <ul><li>Suppresses messages in an area</li><li>Packet loss and information fabrication</li><li>Launch other active attacks</li></ul> | High |